\documentclass[epjCONF,onecolumn]{svjour}
\usepackage{graphics}
\usepackage[varg]{txfonts} 
\usepackage[latin1]{inputenc}
\session-title{New Technologies for Probing the Diversity of Brown Dwarfs and Exoplanets}
\begin{document}
\title{The Calan-Hertfordshire Extrasolar Planet Search}
\author{J.S. Jenkins\inst{1}\fnmsep\thanks{\email{jjenkins@das.uchile.cl}}
  \and H.R.A. Jones\inst{2} \and K. Gozdziewski\inst{3} \and C. Migaszewski\inst{3} \and
J.R. Barnes\inst{2} \and M.I. Jones\inst{1} \and P. Rojo\inst{1} \and
D.J. Pinfield\inst{2} \and A.C. Day-Jones\inst{1} \and S. Hoyer\inst{1}}
\institute{Departamento de Astronomo, Universidad de Chile, Camino el Observatorio 1515, Las
  Condes, Santiago, Chile Casilla 36-D \and Centre for Astrophysics, University of
  Hertfordshire, College Lane, Hatfield, Hertfordshire, AL10 9AB \and
  Torun Centre for Astronomy, Nicolaus Copernicus University, Gagarina 11, 87-100 Torun, Poland}
\abstract{
The detailed study of the exoplanetary systems HD189733 and HD209458
has given rise to a wealth of exciting information on the physics of
exoplanetary atmospheres. To further our understanding of the
make-up and processes within these atmospheres we require a
larger sample of bright transiting planets. We have began a project to 
detect more bright transiting planets in
the southern hemisphere by utilising precision radial-velocity measurements.  We
have observed a constrained sample of bright,
inactive and metal-rich stars using the HARPS instrument and here we 
present the current status of this project, along with our first
discoveries which include a brown dwarf/extreme-Jovian exoplanet found in
the brown dwarf desert region around the star HD191760 and improved
orbits for three other exoplanetary systems HD48265, HD143361 and
HD154672.  Finally, we briefly discuss the future of this project and the
current prospects we have for discovering more bright transiting
planets.
} 
\maketitle
\section{Introduction}
\label{intro}

Since the discovery of the extrasolar planets (aka. exoplanets) 51~Pegasi~$b$ (\cite{Ref1}) and 70~Virginis~$b$ (\cite{Ref2}), arguably 
the most important exoplanetary discoveries were the detections of both HD209458~$b$ and HD189733~$b$.  These planets are the brightest 
transiting exoplanets known to date and both were first detected by radial-velocity variations (\cite{Ref3}, \cite{Ref4}). Since these planets orbit bright stars, 
where radial-velocity samples are 
biased due to the required S/N levels needed for precision work, their transits have allowed follow-up from ground based and space based observatories. 
A number of pivotal discoveries have been made from these systems. For instance, \cite{Ref5} report the detection of both 
oxygen and carbon in the extended upper atmosphere of the planet orbiting HD209458~$b$ using STIS on the Hubble Space Telescope (HST) and \cite{Ref6} also 
used HST data to detect sodium absorption in the planet's atmosphere. The mid infrared 
spectrum of the planet was measured by \cite{Ref7} using the Spitzer IRS instrument, indicating both thermal emission and a possible silicate cloud deck feature at 9.65$\mu$m. 
However, follow-up from \cite{Ref8} has shown that the spectrum is dominated by thermal emission and there is no evidence in their 
data for the silicate feature reported by \cite{Ref7}. Also these data indicate possible heat redistribution between the planet's dayside and 
nightside. In addition, the planet around HD189733 has also been extensively studied. The secondary eclipse 
has been measured (\cite{Ref9}) and has revealed strong infrared emission from this planet. \cite{Ref10} 
claim the discovery of polarized light from this planet, representing the first such detection for any exoplanet. Also, a ground based transmission 
spectrum has revealed sodium absorption (\cite{Ref11}), using the Hobby-Eberly Telescope to observe the planet across 11 orbits, and the 
level was found to be three times larger than that previously found for HD209458~$b$ (\cite{Ref6}). Both \cite{Ref12} and \cite{Ref13} have 
used the HST NICMOS spectrograph to measure water absorption and methane absorption in the atmosphere of HD187933~$b$, along with other molecular bands such as 
carbon monoxide and carbon dioxide, which are the first 
detections of carbon based molecules in any exoplanet. These observations now allow one to compare the 
atmospheric properties of extrasolar planets and also to test numerous model atmospheres (e.g. \cite{Ref14}, \cite{Ref15}, \cite{Ref16}).\newline

\section{The Calan-Hertfordshire Extrasolar Planet Search}
\label{cheps}

From the list of detailed studies shown above, it is clear that a larger census of transiting planets around bright stars will 
provide the exoplanetary community with a database that can be used to examine a wider range of properties that constitute the atmospheres 
of exoplanets. This particularly holds true in the southern hemisphere where no really bright transiting planets yet exist and where some 
of the worlds leading telescopes and instrumentation (e.g. the VLT) lie in wait to examine such systems. Therefore, we have began a targeted 
project in the southern hemisphere that aims to detect more 51~Peg~$b$-like planets and hence more bright transiting exoplanets. The 
Calan-Hertfordshire Extrasolar Planet Search (aka. CHEPS) utilises the radial-velocity technique to monitor a number of bright, inactive and 
metal-rich, solar-type stars to hunt for the typical high frequency radial-velocity amplitude induced by an orbiting short period world.

The current CHEPS target list is drawn from an initial southern sample of $\sim$350 stars that were observed with the ESO-FEROS spectrograph (\cite{Ref17}). The results from 
the analysis of this initial sample were published in \cite{Ref18}. In brief, these stars were all pre-selected from the Hipparcos catalogue to have $B-V$ colours in the range 
0.5$-$0.9 and $V$ magnitudes in the range 7.5$-$9.5. This helps to select solar-type stars that are not on any previous planet search list (i.e. essentially all Sun-like stars 
brighter than 7.5 in $V$ already constitute other planet search target lists) but the stars are bright enough that a detected transiting planet, exhibiting a fairly deep transit light curve, will 
attain benchmark status and provide an ideal target for atmospheric follow-up studies. The FEROS chromospheric activities were then measured following the prescription 
explained in (\cite{Ref19}) to extract each star's log$R'_{\rm{HK}}$-index, which can be used as a proxy for the expected level of radial-velocity jitter (e.g. \cite{Ref20}). 
All stars with log$R'_{\rm{HK}}$ $\le$ -4.5 were included in the CHEPS target list, with the highest priority given to those stars closer to solar activity levels or less (i.e. $\le$-4.9).  
In addition to the activity criteria we also aim to increase the probability of gas giant planet detection by measuring the metallicity, or iron abundance ([Fe/H]) of each of our possible 
planet search targets, since \cite{Ref22} have shown that the probability of gas giant planet detection increases exponentially with the amount of iron found in the photospheres of 
solar-type stars. This result can be explained in the framework of giant planet formation through core accretion of gas depleted materials left over from 
the formation of the parent star (e.g. \cite{Ref24}). Therefore, we only consider stars with a FEROS iron abundance ratio $\ge$+0.1~dex for our planet search project, which 
should help to bias the sample towards more gas giant-like exoplanets. In all, 100 stars were drawn from this pilot sample and almost all have been followed up with 
HARPS to hunt for short period planets.

\begin{figure}
\vspace{5.0cm}
\hspace{-4.0cm}
\includegraphics{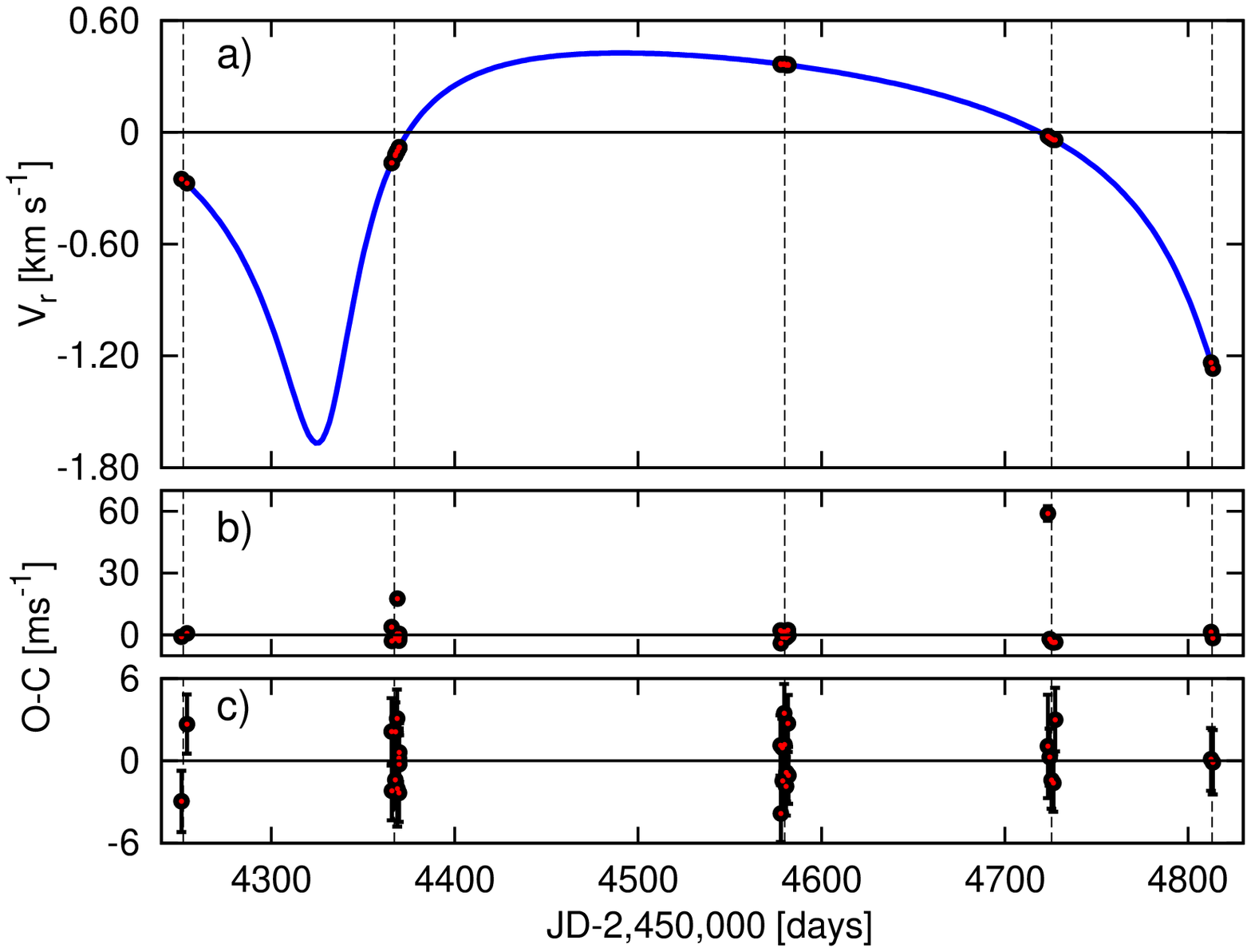}
\includegraphics{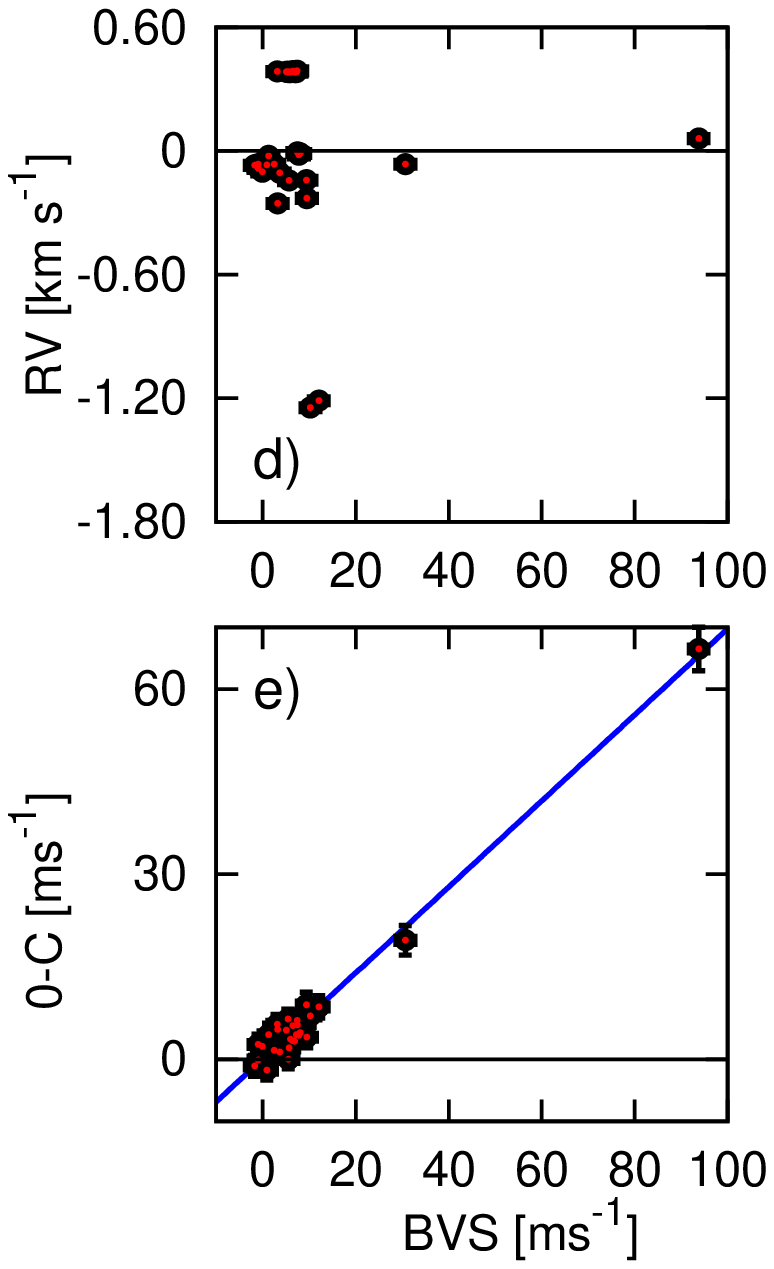}
\vspace{0.2cm}
\caption{Panel (a) shows the best fit Keplerian solution to the Dopper points for HD191760 with BVS correction as in \cite{Ref31}. Panel (b) shows the residuals without BVS correction 
and panel (c) shows the residuals after correction, corresponding to the synthetic curve in panel (a).  Panel (d) shows the radial-velocities against the BVS after subtraction of the 
best-fitting Keplerian synthetic curve. Panel (e) shows the best-fitting linear correlation coefficient of BVS-RV. Note that the two values with the largest BVS help to better constrain the 
correlation.}
\label{fig1}       
\end{figure}

Thus far the CHEPS has obtained at least three radial-velocity points for over 95\% of this pilot sample using the ESO-HARPS instrument (\cite{Ref25}). Of these stars, 
a total of over 25 significant radial-velocity variations have been 
detected (after ruling out spectroscopic binary companions) giving rise to a high fraction of planet like signals ($>$25\%). In \cite{Ref27} we present the first 
results from this data, announcing the discovery of a brown dwarf/extreme-Jovian exoplanet located in the brown dwarf desert region around the star HD191760 (Fig.~\ref{fig1}). 
The likelihood is that this companion is a sub-stellar brown dwarf yet the possibility exists that it is actually a deuterium burning extreme-Jovian exoplanet that formed through 
core accretion. Theoretically such companions are possible up to masses of around 40M$_{\rm{J}}$ orbiting solar-type stars with super-solar metallicities and are likely to be found 
within the brown dwarf desert region at semimajor axes of 1-4~AU (\cite{Ref28},\cite{Ref30}). Panel 
(a) shows all the data obtained for this star (red filled points) $after$ performing a bisector velocity span (BVS) correction (see \cite{Ref31} for details). The solid blue curve reveals a 
companion with a period of 505.65$\pm$0.42~days, a M~sin~$i$ of 38.17$\pm$1.02M$_{\rm{J}}$ and a fairly eccentric orbit of 0.63$\pm$0.01. All orbital parameters are listed in 
\cite{Ref27}. Panels (b) and (c) show the residuals of 
the best fits to the data before and after the BVS correction respectively. Clearly by measuring the BVS and using it to clean the radial-velocity points from any line broadening effects 
such as stellar activity (panels d and e), a much improved solution can be extracted. Also, we ran extensive GAMP stability simulations (\cite{Ref32}) to hunt for regions where 
additional planetary 
mass companions could reside in the system. From this analysis we found that all regions beyond $\sim$0.17~AU and out as far as the detected companion are chaotic, due to the 
disastrous gravity imposed by this massive companion along its eccentric orbit, and therefore if any additional companions reside in the systems within the orbit of this detected 
companion they are likely very low-mass, short period bodies.

\begin{figure}
\vspace{0.cm}
\hspace{-4.5cm}
\includegraphics{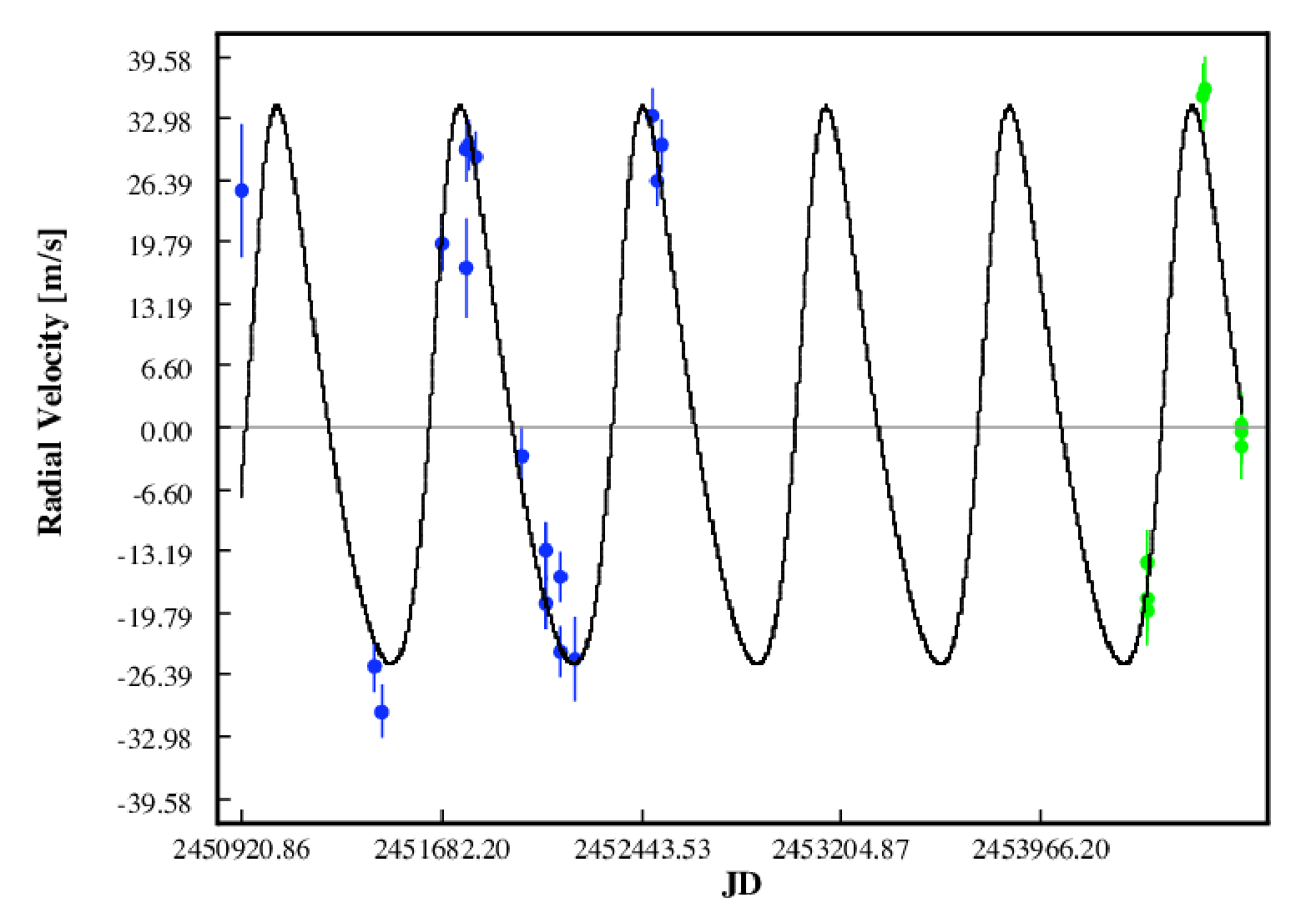}
\includegraphics{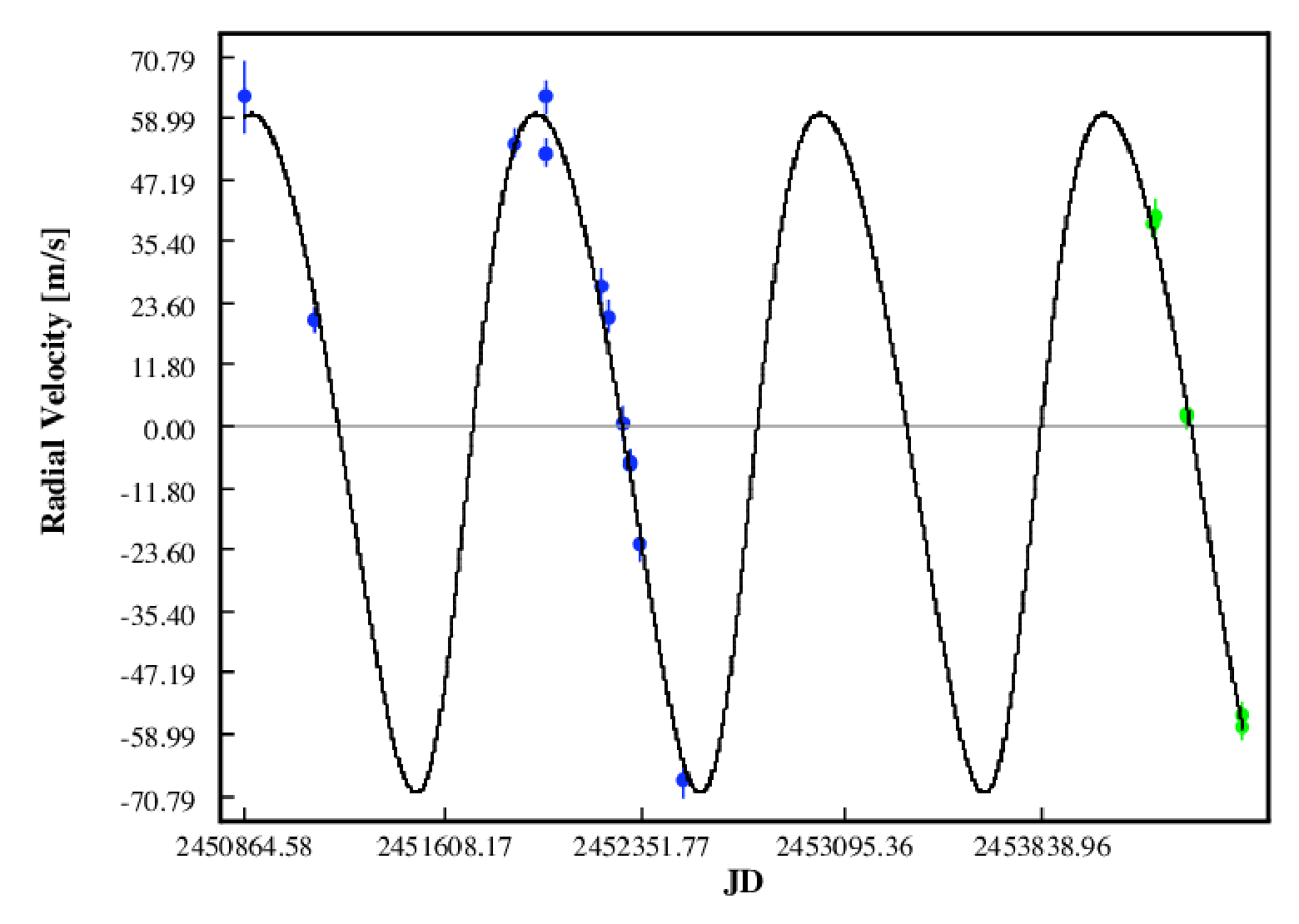}
\includegraphics{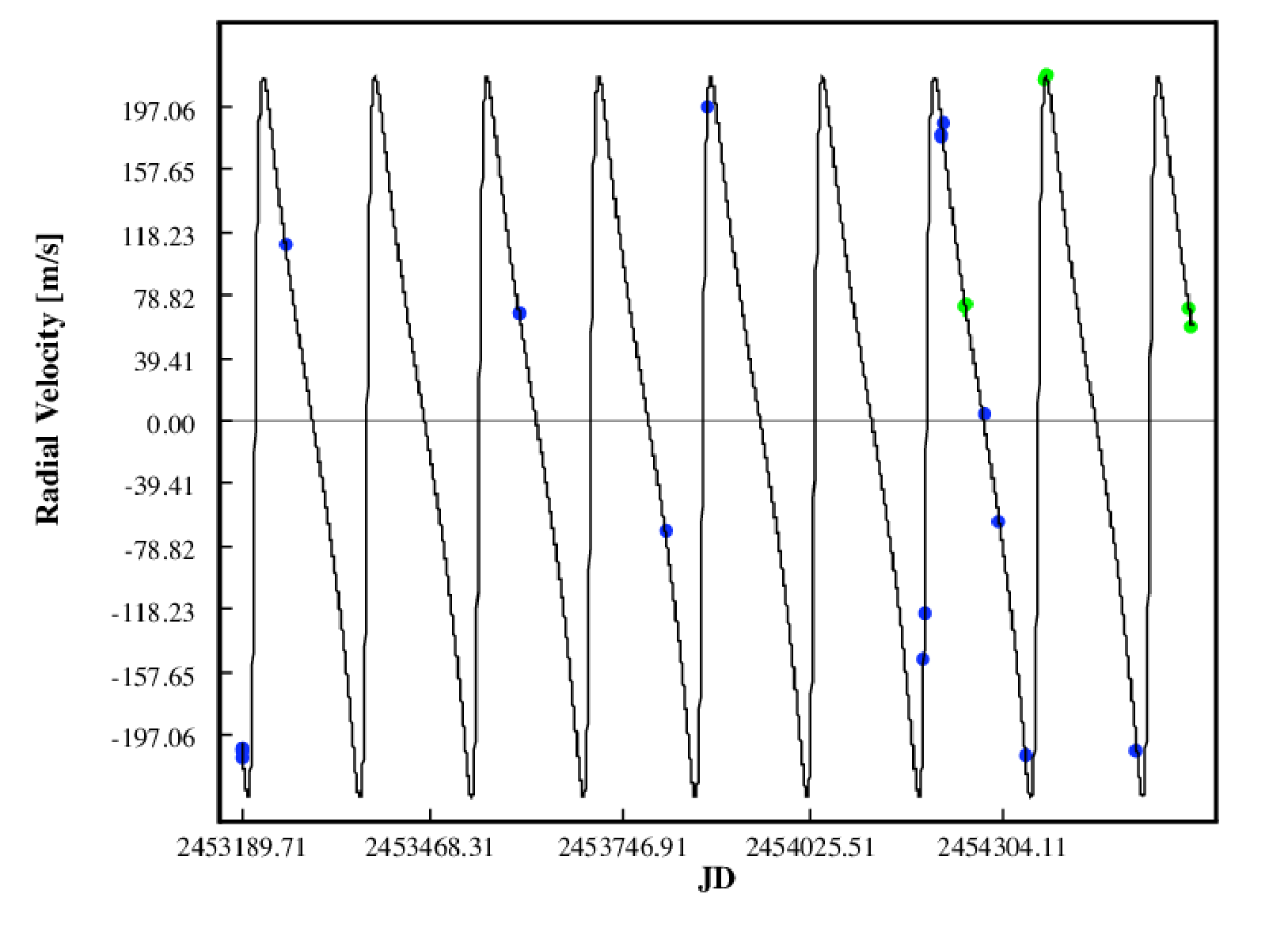}
\vspace{3.3cm}
\caption{The radial-velocity Keplerian fits to the stars HD48265 (left), HD143361 (middle) and HD154672 (right). Plotted in blue are 
the literature values from the Magellan program and in green are the data points from \cite{Ref27}.}
\label{fig2}       
\end{figure}

Along with this interesting discovery, we also show updated orbits for three exoplanets recently announced from the Magellan Planet Search (Fig.~\ref{fig2}). The left panel shows the 
best fit to the star HD48265, the middle panel is for HD143361 and the right panel shows the fit to the data for HD154672, all using the Systemic fitting tool (\cite{Ref33}). The blue 
data points represent the values published by the Magellan Planet Search (\cite{Ref34}, \cite{Ref35}) and the green data are those from our work. The updated orbits for these systems 
are listed in \cite{Ref27} and due to the pre-selection of the target list to focus on the most 
inactive targets, the rms residuals for three of these systems is only $\sim$2-3ms$^{-1}$. HD48265 presents a larger rms scatter due to one errant Magellan data point that has larger 
residuals than the other points and when removed the rms drops to almost 3ms$^{-1}$. All three of these planets are found to have minimum masses that would place them in the gas 
giant regime and none are found to have very short periods. Note that at present we do not find any significant evidence for additional planets in any of these systems.  

Although we initially predicted a yield of 12 planets, as mentioned above, we currently show over 25 planetary-like signals in the pilot sample. This is because the \cite{Ref22} relation 
is only complete for gas giant planets down to minimum masses of 0.4M$_{\rm{J}}$ and with amplitudes ($K$) of $>$30ms$^{-1}$, whereas within our planet-like signals we have a 
number of potential lower mass systems across a wide range of periods.  Such systems require both time and data to properly confirm and constrain.

\section{Summary and Future Work}
\label{summary}

We briefly discuss recent results from the Calan-Hertfordshire Extrasolar Planet Search (CHEPS) that reveal 1) the discovery of a brown dwarf/extreme-Jovian exoplanet in the brown dwarf 
desert region around the star HD191760 and 2) updated orbits for three recently discovered exoplanets (HD48265~$b$, HD143361~$b$ and HD154672~$b$). The CHEPS 
project is currently ongoing and with future time already awarded we expect to announce many more exoplanets around these stars, in addition to more $bright$ 
transiting planets that can be followed up from the ground and space to probe the physics of exoplanetary atmospheres. Finally, we have significantly increased the target sample 
from the initial 100 pilot stars that were drawn from the $\sim$350 stars analysed using FEROS.  We observed a further $\sim$500 stars with FEROS and have recently measured their 
activities and metallicities, giving rise to a three-fold increase in the number of inactive and metal-rich targets we can follow up to hunt for more southern short period exoplanets.

\end{document}